\documentclass[modern]{aastex631}
\usepackage{natbib}
\usepackage{hyperref}
\usepackage{amsmath}
\usepackage{amssymb}
\usepackage{bm}
\usepackage{subfigure}
\usepackage{xcolor}

\NewPageAfterKeywords
\begin{document}

\title{The effect of spatially-varying collision frequency on the development of the Rayleigh-Taylor instability}
\date{November 2023}
\author[0000-0002-7865-5758]{John Rodman}
\affiliation{Virginia Tech, Blacksburg, Virginia 24061}

\author[0000-0001-6835-273X]{James Juno}
\affiliation{Princeton Plasma Physics Laboratory, Princeton, New Jersey 08543}

\author[0000-0002-0712-1097]{Bhuvana Srinivasan}
\email{srinbhu@uw.edu}
\affiliation{Virginia Tech, Blacksburg, Virginia 24061}
\affiliation{University of Washington, Seattle, Washington 98195}

\begin{abstract}
    The Rayleigh-Taylor (RT) instability is ubiquitously observed, yet has traditionally been studied using ideal fluid models. Collisionality can vary strongly across the fluid interface, and previous work demonstrates the necessity of kinetic models to completely capture dynamics in certain collisional regimes. Where previous kinetic simulations used spatially- and temporally-constant collision frequency, this work presents 5-dimensional (two spatial, three velocity dimensions) continuum-kinetic simulations of the RT instability using a more realistic spatially-varying collision frequency. Three cases of collisional variation are explored for two Atwood numbers: low to intermediate, intermediate to high, and low to high. The low to intermediate case exhibits no RT instability growth, while the intermediate to high case is similar to a fluid limit kinetic case with interface widening biased towards the lower collisionality region. A novel contribution of this work is the low to high collisionality case that shows significantly altered instability growth through upward movement of the interface and damped spike growth due to increased free-streaming particle diffusion in the lower region. Contributions to the energy-flux from the non-Maxwellian portions of the distribution function are not accessible to fluid models and are greatest in magnitude in the spike and regions of low collisionality. Increasing the Atwood number results in greater RT instability growth and reduced upward interface movement. Deviation of the distribution function from Maxwellian is inversely proportional to collision frequency and concentrated around the fluid interface. The linear phase of RT instability growth is well-described by theoretical linear growth rates accounting for viscosity and diffusion.
\end{abstract}

\section{Introduction}
Rayleigh-Taylor (RT) instabilities occur when a dense fluid is accelerated into a lighter one, for example under the influence of gravity~\citep{rayleigh1882investigation,taylor1950instability}. 
Traditionally, RT instabilities are studied using fluid models~\citep{ramaprabhu2006limits,sharp1984overview}, yet fully-kinetic simulations demonstrate the existence of regimes of finite collisionality that are RT unstable and exhibit significantly altered dynamics as compared to ideal fluid results~\citep{rodman2022kinetic,sagert2015knudsen}.
Previous simulations investigate the role of kinetic effects such as viscosity, resistivity, and thermal conductivity on the growth of the RT instability \citep{srinivasan2014mitigating,song2020rayleigh,bera2022effect}, but studies of these effects with a fully-kinetic model with a full nonlinear Fokker-Planck collision operator have yet to be performed.

The RT instability appears in high-energy-density regimes such as pulsar wind nebulae like the Crab nebula as a source of large-scale structure and mixing at the surface of the supernova shell~\citep{chevalier1975outer,porth2013threedimensional}, laser implosions at the OMEGA laser facility~\citep{boehly1997omega, smalyuk2009rayleigh, knauer2000single}, and early stages of supernova explosions~\citep{bethe1990supernova, chevalier1978explosions}.
Supernova remnants expanding into the interstellar medium can also give rise to the Rayleigh-Taylor instability~\citep{gull1975xray, chevalier1977interaction}.
Low collisionality and a weak magnetic field serve to reduce classical transport coefficients in the intergalactic medium~\citep{zhuravlena2019suppressed}, so kinetic simulations may be warranted to completely capture interactions of the supernova shell and the interstellar medium.
In the Crab nebula, the interaction between the wind accelerated by the pulsar and the cold supernova shell is an RT-unstable configuration with orders of magnitude of variation in density and pressure across the interface~\citep{jun1998interaction,porth2014rayleightaylor}.
In general, collision frequency is a function of density and temperature~\citep{braginskii1965transport}. 
Densities and temperatures can vary greatly in astrophysical regimes that are RT unstable~\citep{chevalier1977interaction,porth2013threedimensional,gull1975xray}, so collision frequencies are expected to vary similarly. 
Previous fully-kinetic RT instability simulations assumed spatially and temporally constant collision frequency~\citep{rodman2022kinetic}.
This work explores a continuum-kinetic, neutral species RT instability simulations with spatially varying collisionality.
It is the purpose of this work to explore a situation where the collision frequency varies strongly across the interface.

The rest of this paper is organized as follows.
Section \ref{sec:problem} details the governing equation and initial conditions for these simulations.
Results for three cases of collisional variation across the interface are presented in Section \ref{sec:collisions} for an Atwood number of 1/3.
The impact of increasing the Atwood number to 2/3 is discussed in Section \ref{sec:atwood}, and results are compared to the lower Atwood number cases, including growth rates and magnitudes of interface widening.
An expansion of the particle energy-flux is presented in Section \ref{sec:energy}, leveraging the information contained in the distribution function to quantify the importance of kinetic models in these collisional regime.
Finally, Section \ref{sec:summary} summarizes all simulation results and discusses the necessity of kinetic models to accurately model the RT instability in these conditions.

\section{Problem Description}
\label{sec:problem}
Simulations in this work are performed using the continuum-kinetic capabilities of the plasma simulation framework \texttt{Gkeyll} \citep{gkylDocs}. 
\texttt{Gkeyll} utilizes a discontinuous Galerkin discretization scheme \citep{reed1973triangular, cockburn1998runge, cockburn2001runge} with serendipity basis \citep{arnold2011serendipity} to evolve the Boltzmann equation \citep{juno2016discontinuous,hakim2020alias},
\begin{equation}
	\frac{\partial{}f}{\partial t} +  \boldsymbol{v}\cdot\boldsymbol{\nabla}_{\boldsymbol{x}} f + \boldsymbol{g}\cdot\boldsymbol{\nabla}_{\boldsymbol{v}}f = \left( \frac{\partial f}{\partial t}  \right)_C,
\end{equation}
where $f = f(\boldsymbol{x}, \boldsymbol{v}, t)$ is the particle distribution function defined in phase space, $\boldsymbol{g}$ is acceleration due to gravity, and the right-hand-side is the collision operator. 
Where a traditional fluid model assumes particles always follow a thermalized Maxwellian velocity distribution, the continuum-kinetic model allows the velocity space distribution function $f$ to deviate from Maxwellian.
The collision operator relaxes the distribution function to Maxwellian and contains much of the physics that must be explicitly included in fluid models, like viscosity and thermal conduction.
While a full nonlinear Fokker-Planck collision operator \citep{rosenbluth1957fokker} is required to accurately capture the physics of small-angle collisions between charged species, including collision-induced velocity space advection and diffusion and collision frequency that varies in velocity space as $1/v^3$, reduced collision models can be constructed to retain features relevant to the chosen problem.
For example, the Dougherty or Lenard-Bernstein operator \citep{hakim2020conservative}, explicitly includes velocity space advection and diffusion of the distribution function but utilizes a collision frequency that is constant in velocity space, overestimating the impact of collisions in the high-energy tail of the distribution. 
Collisions are modeled in this work by the Bhatnagar-Gross-Krook (BGK) operator \citep{bhatnagar1954model},
\begin{equation}
    \left( \frac{\partial f}{\partial t} \right)_C = \nu (f_M - f),
\end{equation}
where $\nu$ is the collision frequency and $f_M$ is a Maxwellian distribution constructed from moments of $f$.
Where previous work utilized the BGK model with a collision frequency that was constant spatially, collision frequency of a single species generally varies spatially with number density and thermal velocity as \citep{braginskii1965transport},
\begin{equation}
    \nu \propto \frac{n}{v_{th}^{3}}.
\end{equation}

The BGK operator is well-suited to large-angle binary collisions between neutral species and is guaranteed to conserve particle number density, momentum, and energy when collision frequency is constant in velocity space.
However, similarly to the Dougherty operator, the use of mean collision frequency results in an overestimation of collision frequency in the high-energy tail of the distribution.

Simulations in this work are 5-dimensional, with 2 physical space dimensions and 3 velocity space dimensions.
Initial conditions are derived from hydrostatic equilibrium,
\begin{equation}
    \boldsymbol{\nabla} p = -n m \boldsymbol{g},
\end{equation}
where $p$ is pressure, $n$ is number density, $m = 1.0$ is mass, and $g = 1.0$ is gravitational acceleration. Initial number density and pressure profiles are,
\begin{equation}
	n(y) = \frac{n_0}{2} \tanh\left(\frac{\alpha y}{L_y}\right)+ \frac{3}{2}n_0,
\end{equation}
\begin{equation}
	p(y) = -\frac{mgn_0}{2}\left[ \frac{L_y}{\alpha} \ln \left(\cosh\left(\frac{\alpha y}{L_y}\right)\right) + 3y \right] + \frac{3}{2} n_0 T_0,\\
\end{equation}
where $L_y = 1.0$ is half the domain length in $y$, $n_0$ and $n_1$ are the number density at the bottom and top of the domain respectively, and $T_0$ is an arbitrary integral constant chosen such that temperature and pressure remain positive in the domain. 
The width of the density gradient at the center of the domain is proportional to the constant $\alpha$, which is set to 25 to ensure a small interface width relative to the domain size. 
The exact method of determining the bounds of the interface for calculating the growth rate is described in Section 4.
Note that quantities at the lower boundary are denoted with a subscript 0, while those at the upper boundary are denoted with a subscript 1. This initial density profile corresponds to an Atwood number, $\mathit{A} = (n_1 - n_0)/(n_1 + n_0)$, of $1/3$. 
Boundary conditions are periodic in $x$ and static reservoir in $y$, where the boundary cells are a continuation of the initial conditions and do not evolve in time. 
Initial distribution functions are Maxwellian,
\begin{equation}
	f(\boldsymbol{v}) = \frac{n}{(2\pi v_{\mathit{th}}^2)^{3/2}}\exp\left(-\frac{(\boldsymbol{v} - \boldsymbol{u})^2}{2v_{\mathit{th}}^2}\right),
\end{equation}
where $\boldsymbol{u}$ is bulk velocity, and $v_{th} = \sqrt{T/m}$ is thermal velocity with temperature $T = p/n$.

The RT instability is seeded by a single-mode sinusoidal perturbation with wavenumber $k$ applied to the $y$-direction bulk velocity, $u_y$, according to
\begin{equation}
	u_y = -0.1v_{\mathit{th},c}\cos\left(kx\right)\exp\left(-\frac{y^2}{2y_r^2}\right),
\end{equation}
where $k=\pi /(2L_x)$,  $v_{\mathit{th},c}$ is initial thermal velocity at the center of the domain, $L_x = 0.5$ is half the simulation domain length in $x$, and $y_r = L_y/10$ is the characteristic decay length for the perturbation.

For the hydrostatic equilibrium chosen in this work, the collision frequency profile defined in Eq. 3 increases dramatically near the upper boundary due to the temperature gradient, as shown in Figure 1.
\begin{figure}[]
    \centering
    \includegraphics[width=0.6\linewidth]{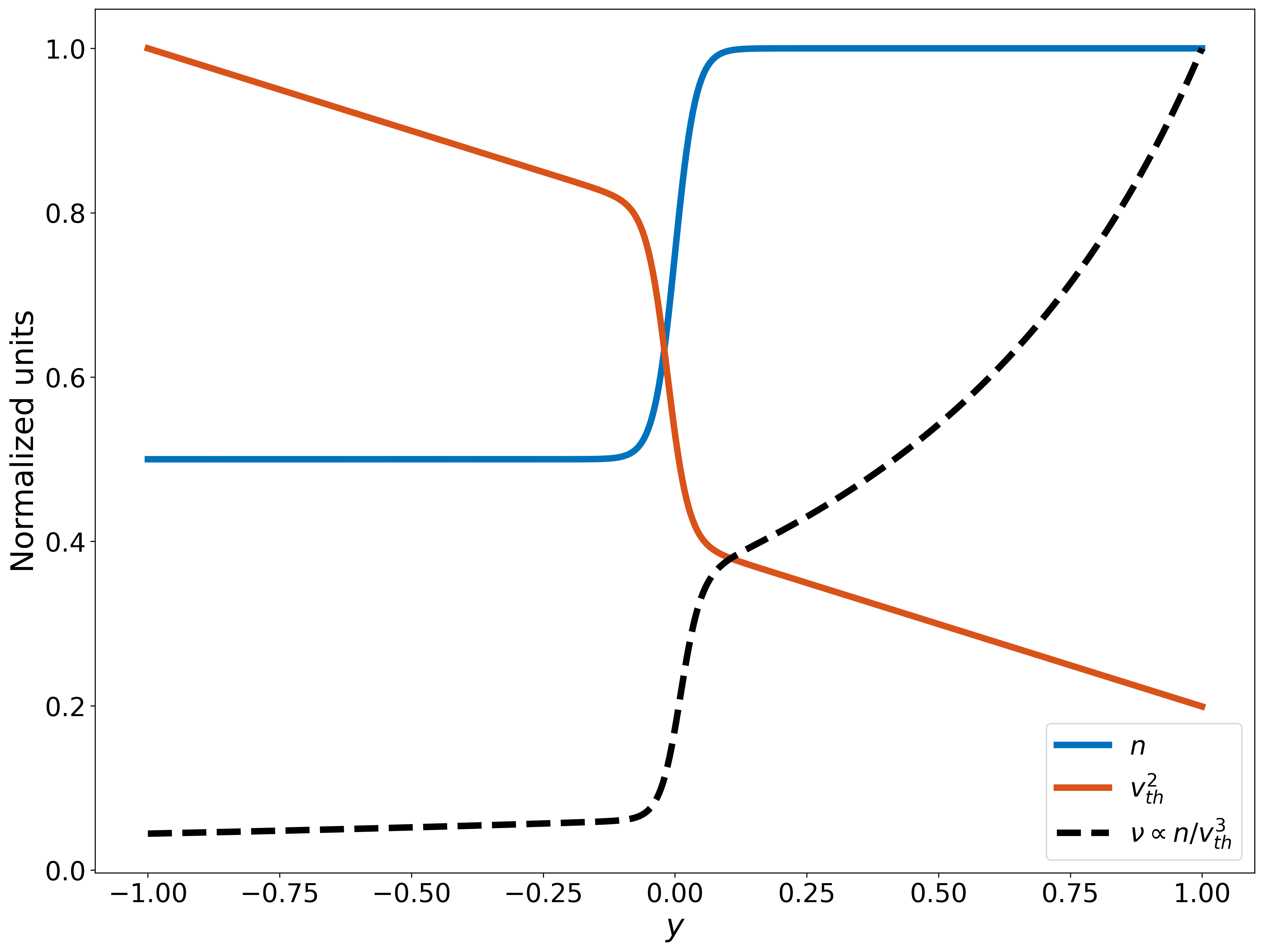}
    \caption{Initial conditions of number density, $n$, thermal velocity squared, $v_{th}^2$, and common collision frequency profile, $n/v_{th}^3$. Note the steep increase in collision frequency approaching the upper boundary. A collision frequency profile proportional to a power of number density maintains similar variation across the interface while avoiding an excessively small time-step for dynamics away from the interface.}
    \label{fig:ICs}
\end{figure}

Therefore, to capture similar collisional variation across the interface while maintaining a reasonable time-step, collision frequency in this work varies according to
\begin{equation}
    \nu = \nu_0 n^\beta,
\end{equation}
where $\nu_0$ is an arbitrary scaling constant and $\beta$ is a parameter chosen to adjust the degree of variation of collisionality across the interface.
As a result of this simplification, collision frequency near the upper boundary will be underestimated by up to a factor of approximately 5. 
This underestimation is acceptable because the high collisionality chosen in this upper boundary is already in a fluid-like regime, and an even higher collisionality would asymptote to a fluid regime without significantly impacting the results at the Rayleigh-Taylor-unstable interface.
Relative collisionality is quantified by the Knudsen number $\mathit{Kn} = \lambda_m/L_x$ where $\lambda_m = v_{th}/\nu$.

\section{Effect of Varying Collisionality}
\label{sec:collisions}
Three cases of varying collisionality are selected to cover regimes previously studied with constant collisionality \citep{rodman2022kinetic}. Table 1 shows the collisional variation across the interface for each case. Note with the same equilibrium profile, the case of constant collisionality with $\mathit{Kn}$ of 0.1 exhibits no RT instability growth, $\mathit{Kn}$ of 0.01 showed diffusion of the interface and limited instability growth, and $\mathit{Kn}$ of 0.001 showed limited interface diffusion and growth similar to an ideal fluid result.

\begin{table}[h!]
    \centering
    \begin{tabular}{|l|l|l|}
    \hline
    Case & Lower $\mathit{Kn}$ & Upper $\mathit{Kn}$ \\ \hline
    1    & 0.1      & 0.01     \\ \hline
    2    & 0.01     & 0.001    \\ \hline
    3    & 0.1      & 0.001    \\ \hline
    \end{tabular}
    \caption{Knudsen numbers defined at the lower and upper ends of the Rayleigh-Taylor interface for each of the three cases.}
\end{table}

\begin{figure}[]
    \centering
    \includegraphics[width=0.6\linewidth]{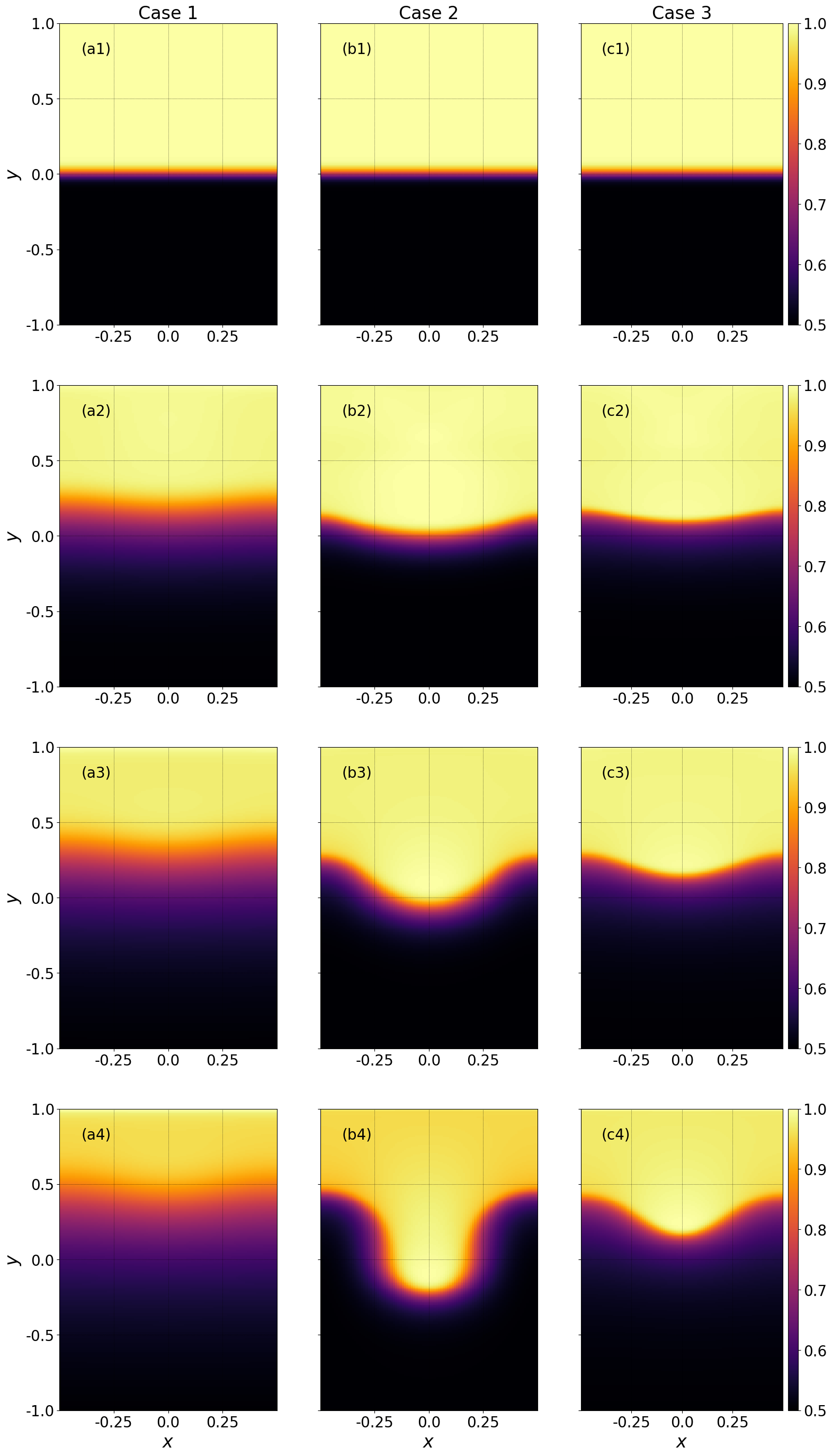}
    \caption{Evolution of number density for three Rayleigh-Taylor instability simulations with Knudsen number varying from: 0.1 to 0.01 (a, left), 0.01 to 0.001 (b, center), 0.1 to 0.001 (c, right). Knudsen numbers are calculated at the lower and upper ends of the interface, respectively. Note the lack of instability growth in case 1, fluid-like growth in case 2, and damped growth in case 3.}
    \label{fig:numDens}
\end{figure}

\begin{figure}[h!]
    \centering
    \includegraphics[width=0.6\linewidth]{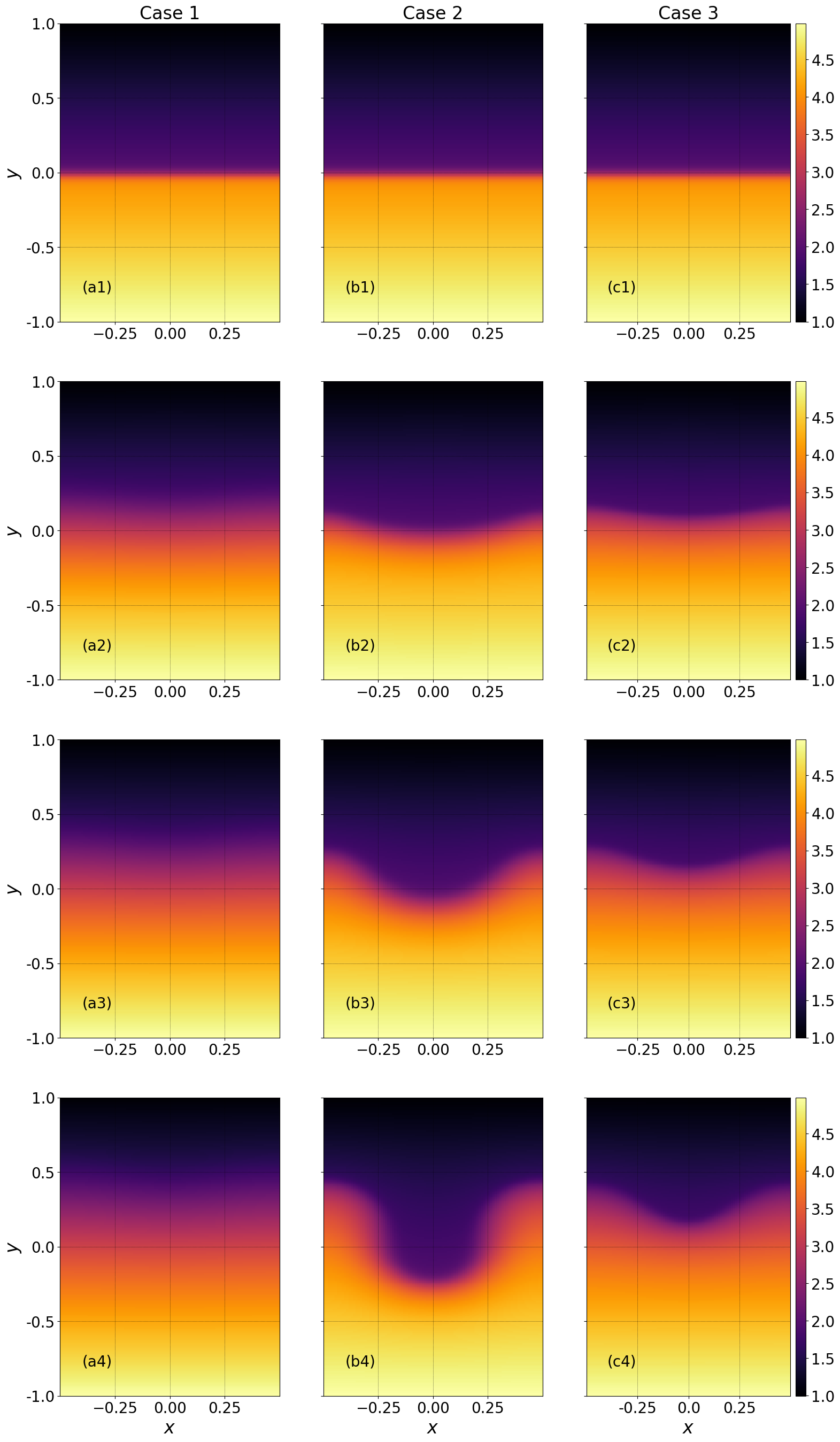}
    \caption{Evolution of $v_{th}^2$ for three Rayleigh-Taylor instability simulations with Knudsen number varying from: 0.1 to 0.01 (a, left), 0.01 to 0.001 (b, center), 0.1 to 0.001 (c, right). Knudsen numbers are calculated at the lower and upper ends of the interface, respectively. Note the magnitudes of $v_{th}^2$ remain stable in time, without the compressive effects seen in Figure \ref{fig:numDens}.}\label{fig:temp}
\end{figure}

Evolution of the number density to the final time of 3 classical RT growth periods, $3\tau_\text{RT} = 3/\sqrt{kgA},$ for each case is presented in Figure~\ref{fig:numDens}. 
Case 1 exhibits no instability growth and is dominated by diffusion of the interface, similar to the constant collisionality case with $\mathit{Kn} = 0.1$.
In case 2, there is early-time diffusion of the interface as the characteristic bubble and spike structures of the RT instability begin to form, but late in time, diffusion appears to be limited as instability growth becomes dominant.
At the end time (b4), there is clear development of the RT instability, and the average center position between the bubble and spike has moved upwards due to diffusion in the lower, less collisional region.
Additionally, the interface has variable width, with the interface appearing thinner at the peaks of the bubble and spike than in the intermediate vertical regions.
As will be discussed in Section \ref{sec:atwood}, interface width varies between the bubble and spike, and Fig. \ref{fig:diffusion} highlights the evolution of interface width in time.
In case 3, the factor of 100 variation in collisionality between regions drives the interface upwards immediately, with diffusion being strongly biased on the lower side of the interface.
Instability growth is greatly limited relative to case 2 as the lower collisionality region damps the growth of the downward spike.
This is most clearly seen comparing the end time number densities of case 2 (b4) and case 3 (c4), where the bubble reaches approximately the same position at $y\approx 0.5$, while the spike in case 3 is well above $y = 0$ compared to $y\approx -0.25$ in case 2.
Variation in interface width in case 3 also appears to be less substantial than in case 2, likely due to the limited instability growth.
Figure~\ref{fig:temp} shows the evolution of temperature with time. 
The temperature distribution is nearly identical to the number density distribution at each time, though compressibility effects are less prevalent due to heat flux, yielding a smoother profile.

\begin{figure}[h!]
    \centering
    \includegraphics[width=0.7\linewidth]{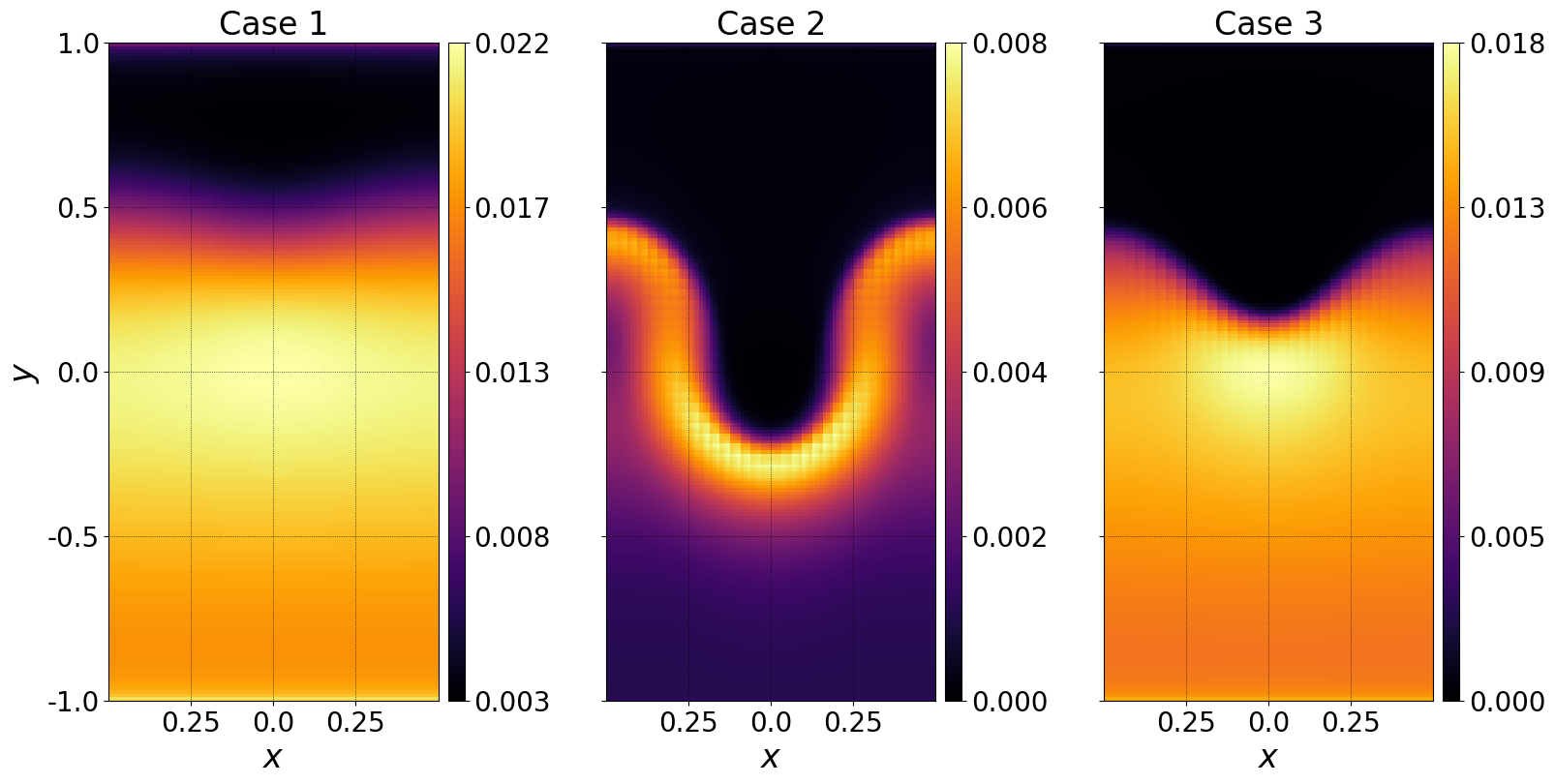}
    \caption{Density of non-Maxwellian distribution normalized to local number density, $n_N/n$, for each case at time $3.0\tau_\text{RT}$. Around the interface in case 2 and in regions of low collision frequency in cases 1 and 3, $n_N/n$ reaches maximum value.}\label{fig:nonMax}
\end{figure}

In a similar manner to \cite{pezzi2021dissipation} and \cite{greco2012imhomogeneous}, non-equilibrium kinetic effects can be quantified using a non-Maxwellian density, $n_N$, constructed from the distribution function and its associated Maxwellian, $f_M$ as,
\begin{equation}
    n_N(\boldsymbol{x}) = \int | f_M(\boldsymbol{x}, \boldsymbol{v}) - f(\boldsymbol{x}, \boldsymbol{v})| 
    \text{d}^3 \boldsymbol{v}.
\end{equation}
This diagnostic has units of density and can be interpreted as the density of non-Maxwellian distribution function. 
Figure~\ref{fig:nonMax} shows the distribution of $n_N/n$, the fraction of non-Maxwellian distribution, at the final time $3.0\tau_{\text{RT}}$.
In each case, $n_N/n$ has higher magnitudes in the regions of lower collisionality, and global magnitudes of $n_N$ are small relative to number density, on the order of 1\%. 
There is correspondence between collision frequency and $n_N/n$ because lower collision frequencies will not be able to thermalize the distribution function as quickly.
The maximum values of $n_N/n$ are similar in cases 1 and 3, as expected due to the lower regions having the same collision frequency.
Similarly, the minimum values of $n_N/n$ approach 0 in cases 2 and 3, as the highly-collisional regions in those cases are similar to an ideal fluid-like regime.
Non-Maxwellian density reaches its maximum around the interface in each case, implying the presence of kinetic effects around the bubble and spike of the RT instability.
The peaks of the bubble and spike do not have equal magnitudes of $n_N/n$; the center of the spike is the absolute maximum of $n_N/n$ in both cases 2 and 3.
This is likely connected to the damping of downward instability growth in the low-collisionality lower region due to diffusion.

\section{Effect of Varying Atwood Number}
\label{sec:atwood}
\begin{figure}[h!]
    \centering
    \includegraphics[width=0.8\linewidth]{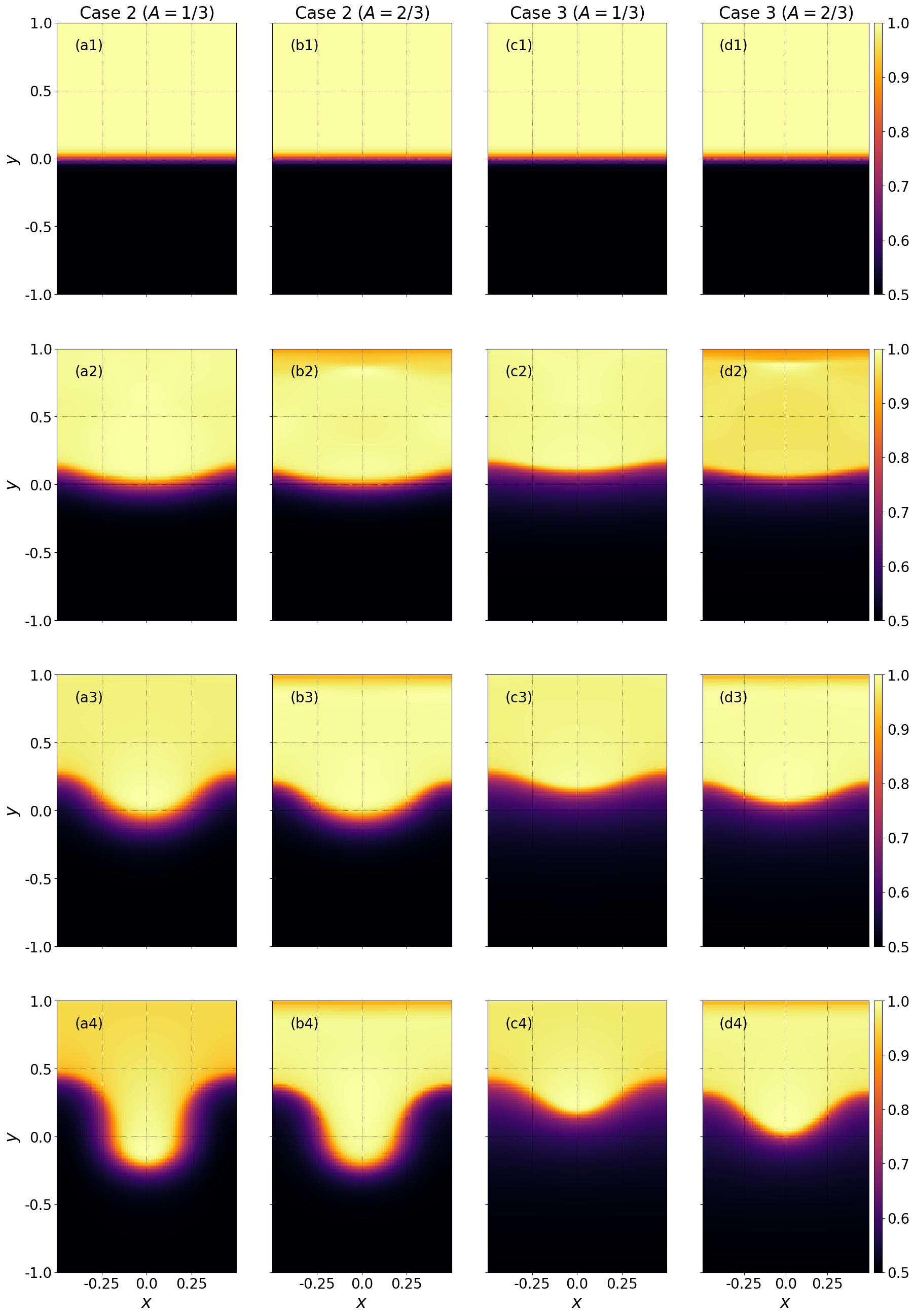}
    \caption{Evolution of number density for Rayleigh-Taylor instability simulations with Knudsen number varying from: 0.01 to 0.001 (left two columns) and 0.1 to 0.001 (right two columns). Knudsen numbers are calculated at the lower and upper ends of the interface, respectively. Results are included for Atwood numbers of 1/3 (a, c) and 2/3 (b, d). Relative to the lower Atwood number cases, in the 2/3 Atwood number cases, the bubble does not move as far upward, but the spike extends further down into the low-density region. There is also less upward movement of the interface due to particle streaming diffusion in the lower, less-collisional region.}\label{fig:numDens23}
\end{figure}

\begin{figure}[h!]
    \centering
    \includegraphics[width=0.5\linewidth]{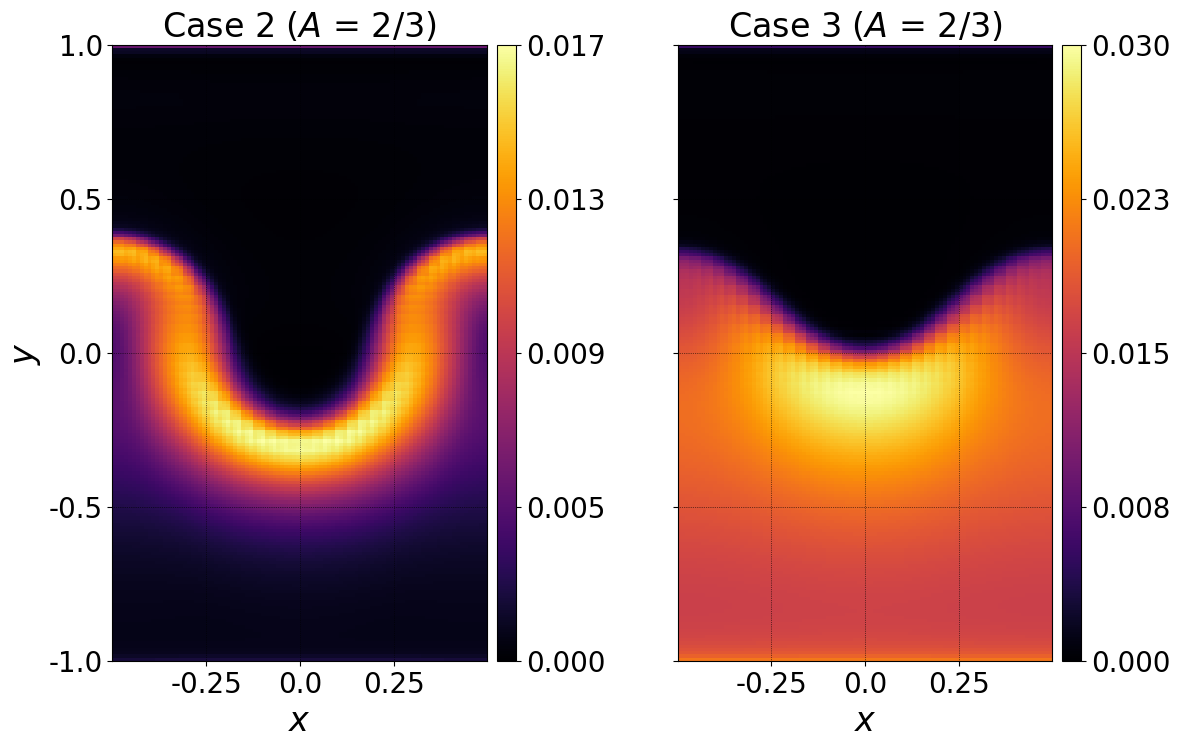}
    \caption{Density of non-Maxwellian distribution normalized to local number density, $n_N/n$, for cases 2 and 3 with Atwood number $A = 2/3$ at time $3.0\tau_\text{RT}$. Similar to the $A=1/3$ cases, the peaks in $n_N/n$ appear at the interface in case 2 and in the lower collision frequency region in case 3.}\label{fig:nonMax23}
\end{figure}

Previous simulations have focused on varying collision frequency with a given equilibrium profile and Atwood number.
Atwood number is known to vary greatly across astrophysically-relevant regimes \citep{ebisuzaki1989rayleigh}, so the effect of a different Atwood number on RT instability growth is worth investigating.
For these simulations, $A$ is increased to 2/3 with a similar equilibrium profile by adjusting equations 6 and 7 while maintaining the same collision profile.
Figure~\ref{fig:numDens23} shows the evolution of number density to the same normalized time $3.0\tau_{\text{RT}}$ for the two cases where the RT instability develops with Atwood number of 2/3.
Note the fluctuations in density early in time are caused by waves launched early in time from the initial perturbation in bulk velocity and appear to be more significant in magnitude relative to those in the lower Atwood number cases.
Relative to the $A = 1/3$ cases, there is less upward movement of the interface due to diffusion, yet formation of the downward spike is still damped.
In both cases, the bubble does not move as far upward as the corresponding $A = 1/3$ cases, but the spike reaches further downward, yielding larger total instability amplitude.
The difference in spike position is especially clear when comparing the case 3 simulations, as the spike in the lower Atwood number case is in the upper half of the domain, while the spike remains around or below $y=0$ in the higher Atwood number case.
Normalized non-Maxwellian density $n_N/n$ for these cases is shown in Figure~\ref{fig:nonMax23}.
Spatial distributions of $n_N/n$ follow similar patterns to the lower $A$ cases.
However, magnitudes of $n_N/n$ for the higher $A$ simulations are approximately twice those of the lower $A$ cases.
This is likely due to the proportionality of the kinematic viscosity and diffusion coefficient to thermal velocity (Eq. 11) and the fact that the higher Atwood number cases are less dense, yielding a higher temperature for the same hydrostatic equilibrium.

Growth of the RT instability can be tracked in time by calculating the amplitude between the center of the interface at the peaks of the bubble and spike.
In order to calculate the amplitude, the $y$-location of a reference density, chosen as the initial value of $n$ at the center of the interface, is determined for each frame at $x=0$ and $x=L_x$ for the spike and bubble, respectively.
As will be discussed at the end of this section, compressibility leads to buildup of density on either side of the interface, but the chosen reference density generally remains at the center of the interface in time.
The vertical displacement between those two locations is taken as the amplitude, $h$.

\begin{figure}[h!]
    \centering
    \includegraphics[width=0.55\linewidth]{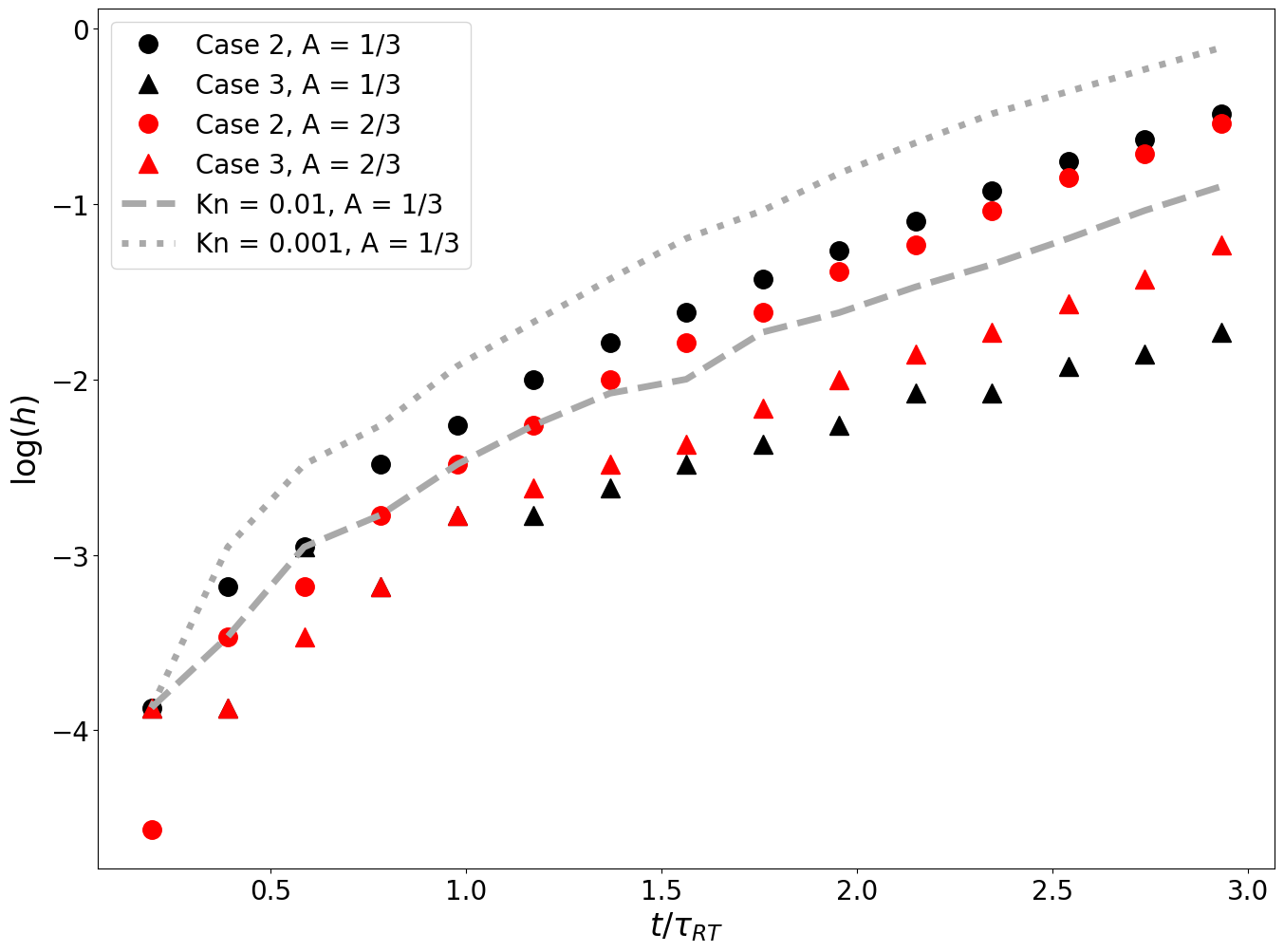}
    \caption{Rayleigh-Taylor amplitude growth. The most fluid-like kinetic case (grey dotted line) has the greatest overall instability growth. Case 2 for both Atwood numbers sits between the fluid-like and intermediate (dashed grey line) cases, while case 3 exhibits the least growth. }\label{fig:growth}
\end{figure}

Figure~\ref{fig:growth} shows the logarithm of the amplitude for cases of constant and varying collisionality that exhibits RT instability growth.
The classical RT growth rate, $\gamma_\text{RT} = 1/ \tau_{\text{RT}} = \sqrt{kgA}$ is not expected to capture the kinetic dynamics included in these simulations, so a growth rate is calculated in a similar manner to \citep{sagert2015knudsen,duff1962effects}, including viscous and diffusive effects, 
\begin{equation}
    \gamma_{0} = \sqrt{kgA+ \nu_v^2k^4} - (\nu_v + \xi)k^2,
\end{equation}
where $\nu_v = v_{\mathit{th},c}\lambda_m/2$ is the kinematic viscosity, and $\xi = \nu_v$ is the diffusion coefficient. 
Dynamic diffusion effects may also be included in this calculation to give a time-dependent growth rate, but this is excluded for this calculation for simplicity. 
Dynamic diffusion dominates early in time, before the linear phase, \citep{luo2020effects} when the interface is diffusing with no instability growth, leading to nonlinear interface amplitude, so early-time points are excluded from the linear fit.
Table 2 shows the growth rates for each case as calculated from the linear fit, $\gamma$, and from equation 11, $\gamma_0$.
\begin{table}[h!]
    \centering
    \begin{tabular}{|l|l|l|l|}
    \hline
    \multicolumn{1}{|c|}{Case} & \multicolumn{1}{c|}{$\gamma$} & \multicolumn{1}{c|}{$\gamma_0$} & \multicolumn{1}{c|}{$\gamma/\gamma_0$} \\ \hline
    Kn = 0.01 (A = 1/3)  & 0.789 & 0.911 & 0.866 \\ \hline
    Kn = 0.001 (A = 1/3) & 0.903 & 1.012 & 0.892 \\ \hline
    Case 2 (A = 1/3)     & 0.879 & 0.989 & 0.889 \\ \hline
    Case 3 (A = 1/3)     & 0.574 & 0.981 & 0.585 \\ \hline
    Case 2 (A = 2/3)     & 1.396 & 1.408 & 0.991 \\ \hline
    Case 3 (A = 2/3)     & 1.130 & 1.373 & 0.823 \\ \hline
    \end{tabular}
    \caption{Growth rates of the Rayleigh-Taylor instability as calculated from linear fit, $\gamma$, and from theory accounting for diffusion and viscosity (Eq. 11), $\gamma_0$.}
\end{table}
Note the classical growth rates are $\gamma_{\text{RT},1} = 1.023$ and $\gamma_{\text{RT},2} = 1.447$ for $A = 1/3$ and $2/3$, respectively.

Theoretical growth rates $\gamma_0$ increase with average collision frequency, approaching the classical growth rates $\gamma_\text{RT,1}$ and $\gamma_\text{RT,2}$.
Regardless of Atwood number, as average collisionality increases, i.e. Case 3 to Case 2 and Kn = 0.01 to Kn = 0.001, the ratio $\gamma/\gamma_0$ increases as the calculated growth rate approaches the theoretical result.
For the cases of spatially-varying collisionality, the degree to which the ratio increases with average collision frequency is dependent on the Atwood number.
The ratio $\gamma/\gamma_0$ for the $A=1/3$ cases increases from 0.585 to 0.889 from Case 3 to Case 2, an increase in agreement of approximately 34\%.
Similarly, the $A=2/3$ cases increase from 0.823 to 0.991, an increase of approximately 17\%.
Therefore, for the same given collisionality profile, the agreement of the calculated growth rate with the theoretical growth rate increases with Atwood number, but as collisionality increases, the relative increase in agreement is greater in the lower Atwood number simulations. 

A primary distinguishing factor between simulations is the magnitude of interface diffusion, which can be quantified by tracking the width of the interface in time.
Traditionally in fluid simulations, the moving interface can easily be tracked in time using fluid mass fraction or by following the constant density values that define the bounds of the interface.
However due to compressibility, the number density profile is not constant around the interface in these simulations, making it difficult to exactly define and track the interface.
Therefore, in an approach similar to \cite{lai2016nonequilibrium}, the interface is instead tracked using the temperature, which remains relatively smooth in time because of heat fluxes smoothing perturbations.
The bounds of the interface are determined from the initial conditions by first retrieving the location where $n$ reaches 99\% of the global maximum as the upper bound and defining that position as $y=L_\text{int}$.
The initial upper and lower bounds of the interface are then located at $y = L_\text{int}$ and $y = -L_\text{int}$, respectively.
Then $v_{th}^2$ is evaluated at each of these points to determine the reference values that are used to track the interface.
For each data frame, the reference values of $v_{th}^2$ are matched at $x=0,L_x$ for the spike and bubble, respectively.

\begin{figure}
    \centering
    \plottwo{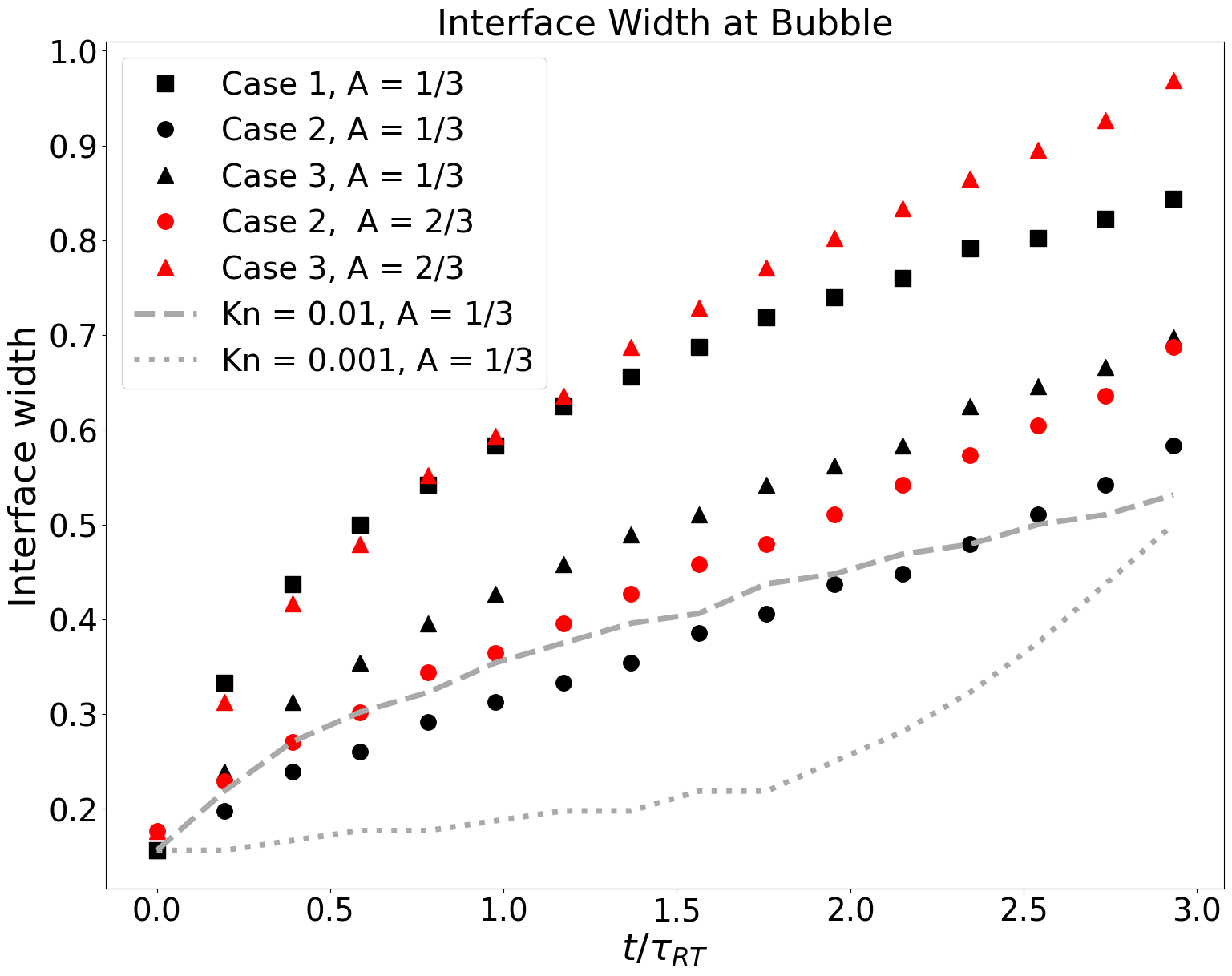}{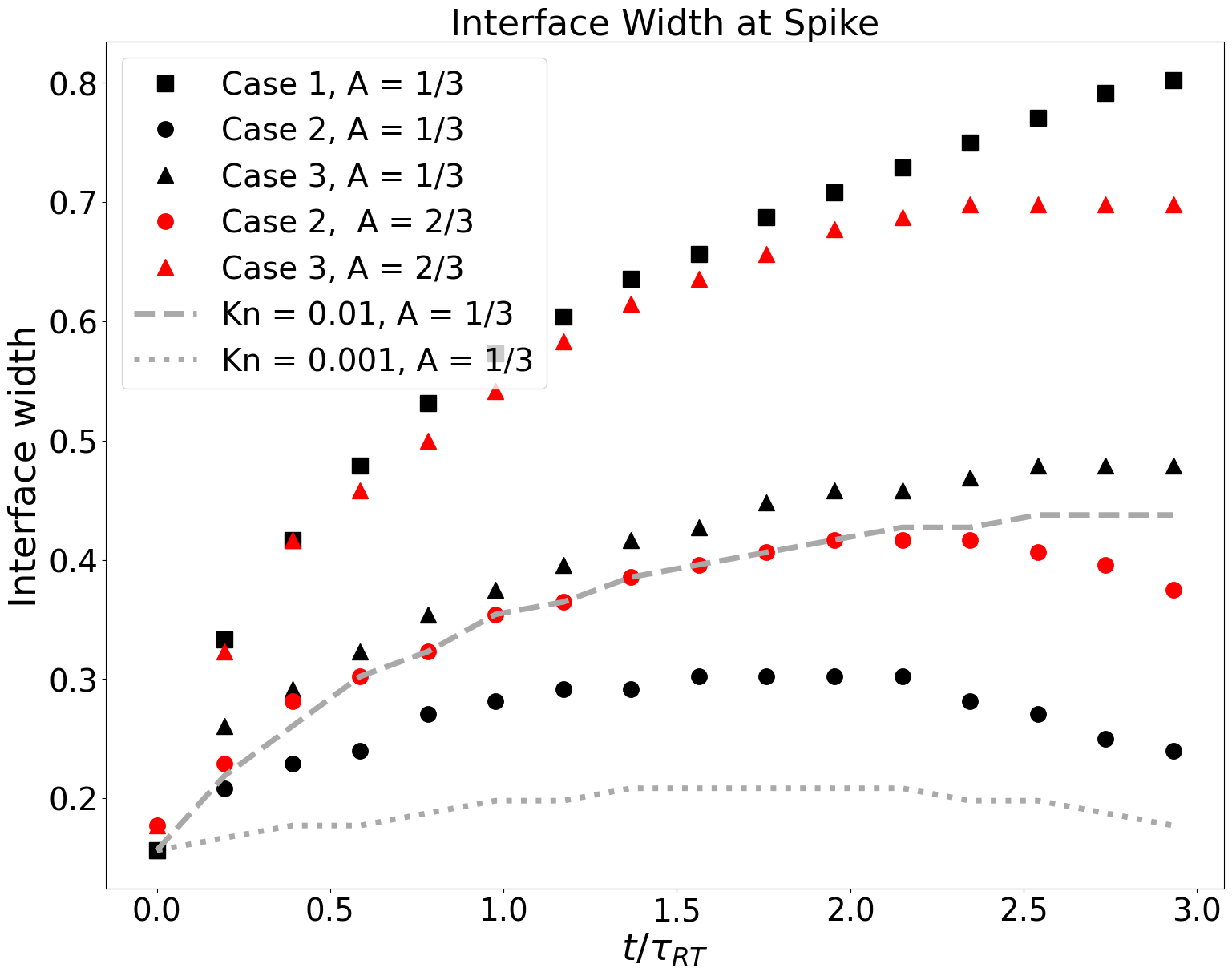}
    \caption{Evolution of the width of the interface between regions as measured at the location of the peak of the bubble (left) and spike (right). Higher Atwood number cases exhibit more interface diffusion than corresponding cases with lower Atwood numbers. The interface width at the bubble is in general larger than at the spike due to diffusion in the low collision frequency region diffusing the lower end of the interface away from the upward moving spike. Note the increase in constant $\mathit{Kn}=0.001$ (grey dotted line) is due to secondary instabilities.}\label{fig:diffusion}
\end{figure}

Figure~\ref{fig:diffusion} shows the evolution of the interface width as calculated at the peak of the bubble and spike.  Differences in interface width between the bubble and spike can be attributed to the difference in collision frequency between the upper and lower regions. 
In the upper region, collision frequency is greater, so it is expected that the upward moving bubble exhibits more ideal fluid-like behavior, i.e.\ faster instability growth and less diffusion. 
Conversely, the lower region with lower collision frequency is dominated by diffusion and damps instability growth, as can be most easily seen in the final number density distributions of case 3.
Therefore, diffusion is in general biased toward the less collisional lower region.
As the bubble and upper end of the interface move upward, the lower end of the interface diffuses downward, resulting in generally larger interface width relative to the spike, which pushes into the lower region as the interface diffuses in the same direction. 
This can be seen by comparing case 1 with $A = 1/3$ (black squares) and case 3 with $A = 2/3$ (red triangles) in Figure \ref{fig:diffusion}, which have the same collisionality in the lower region.
At the bubble location, the interface width of case 3 is increased by the upward movement of the bubble, yielding a wider interface than case 1, which is effectively pure diffusion.
Conversely, the spike in case 3 moves slightly below $y = 0$, somewhat offsetting the interface width gain due to diffusion, so case 1 has a wider interface at that location.
Additionally, in some cases the lower end of the interface diffuses enough that it reaches the lower boundary, so the interface width late in time reaches a maximum. 
If the domain was larger, diffusion would likely continue to follow the early time trend and increase as a similar rate. 

Constant collisionality cases are included as dashed lines in Figure~\ref{fig:diffusion}. 
The most collisional and fluid-like case, constant $\mathit{Kn} = 0.001$, has substantially less interface diffusion than any of the other cases and has effectively a constant interface width in time.
Note the increase at the bubble late in time is an artifact due to the development of secondary instabilities. 
At the bubble, the intermediate constant $\mathit{Kn} = 0.01$ case closely matches case 2 for both Atwood numbers, which have a Knudsen number of 0.01 in the upper region.
However, at the spike, the intermediate constant $\mathit{Kn}$ case still matches the $A=2/3$ case well, while the lower $A$ case deviates from both, exhibiting much less interface widening.
Cases with $A = 2/3$ show greater interface width than the corresponding case with $A = 1/3$, likely due to the higher thermal velocity and therefore diffusion coefficient.
Comparable fluid simulations with varying viscosity exhibit no interface diffusion, similar to the fluid-like $\mathit{Kn}=0.001$ case \citep{bera2022effect,song2020rayleigh,song2020unstructured}.
Viscous fluid simulations also exhibit stabilization of short-wavelength modes, an effect that is not seen in these large single-mode kinetic simulations.
Stabilization in small RT instability modes in kinetic simulations with finite collision frequency is expected to differ from previous fluid simulations, as Braginskii viscosity only applies in the high-collisionality limit.

\section{Expansion of the Particle Energy-Flux}
\label{sec:energy}
\begin{figure}[h!]
    \centering
    \includegraphics[width=0.8\linewidth]{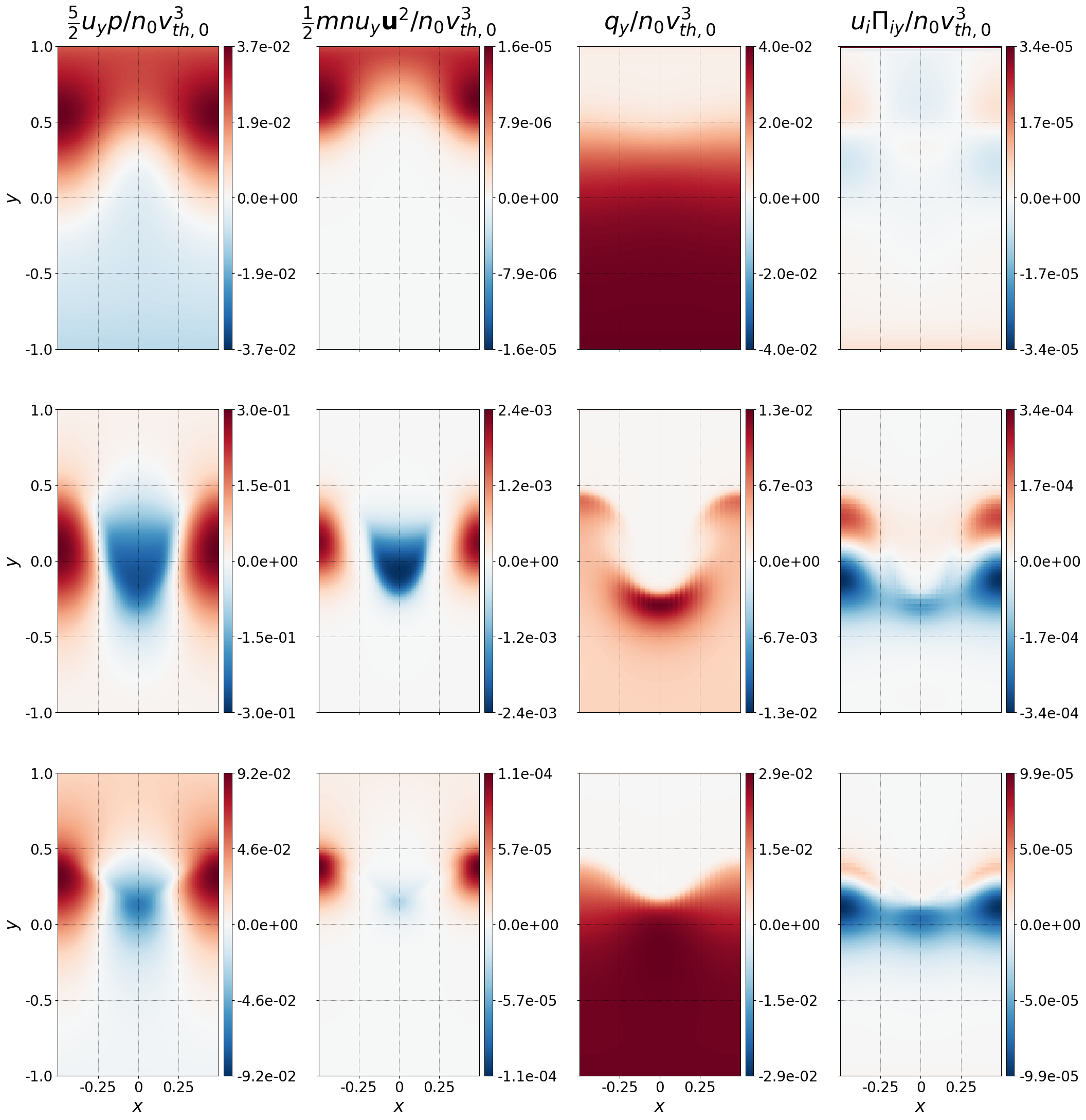}
    \caption{Terms in the expanded particle energy-flux, Eq. 14, for cases 1 (top), 2 (center), and 3 (bottom). Ideal terms (left two columns) are concentrated in the highly-collisional region in case 1 and in the bubble and spike in cases 2 and 3. The non-ideal terms (right two columns) are concentrated in the low-collisional regions in all cases with extrema present around the interface in cases 2 and 3.}\label{fig:energyFlux}
\end{figure}

Higher moments of the distribution function can also be utilized to characterize the impact of kinetic effects. Beginning with the laboratory-frame second and third moment,
\begin{equation}
    \mathcal{P}_{\mathit{ij}} = m\int v_i v_j f \text{d}^3\boldsymbol{v},
\end{equation}
\begin{equation}
	\mathcal{Q}_{\mathit{ijk}} = m\int v_i v_j v_k f\text{d}^3 \boldsymbol{v}.
\end{equation}
As in \citet{wang2015comparison}, by defining $w_i = v_i - u_i$, Eq. (11) can be expanded and tensor contracted to get the particle energy-flux (using Einstein's summation convention),
\begin{equation}
    \frac{1}{2}\mathcal{Q}_{\mathit{iik}} = \underbrace{\frac{5}{2}u_kp + \frac{1}{2}mnu_k\boldsymbol{u}^2}_{\text{I}} + \underbrace{q_k + u_i\Pi_{\mathit{ik}}}_{\text{II}},
\end{equation}
where
\begin{equation}
    q_k = \frac{1}{2}m\int w_iw_iw_kf\text{d}^3\boldsymbol{v},
\end{equation}
is the heat flux vector in the gas frame, and the stress tensor $\Pi_{\mathit{ij}}$ is related to the pressure tensor,
\begin{equation}
    P_{\mathit{ij}} = m\int w_iw_jf\text{d}^3\boldsymbol{v},
\end{equation}
by $\Pi_{\mathit{ij}} = P_{\mathit{ij}} - p\delta_{\mathit{ij}}$ with scalar pressure $p = P_{\mathit{ii}}/3$. 
The pressure tensor is also related to the second moment by $\mathcal{P}_{\mathit{ij}} = P_{\mathit{ij}} + mnu_i u_j$. 
The four terms in the expanded particle energy-flux, Eq. 14, can be grouped into terms that arise from the Maxwellian parts of $f$, group I, and from the non-Maxwellian parts, group II.
Comparing the magnitudes of group I and group II terms quantifies the relative contributions to the total energy-flux of effects that would not be captured by pure fluid models.
Figure~\ref{fig:energyFlux} shows the $y$-components of the expanded energy-flux for cases 1, 2, and 3 for $A = 1/3$, normalized to $n_0 v_{th}^3$. 
Beginning with case 1 (Figure~\ref{fig:energyFlux}, top row), the ideal terms (left two columns) reach maximum values in the more collisional upper region, while the dominant non-ideal term, $q_y$ reaches its maximum in the low-collisional lower region.
All energy-flux terms in case 2 (Figure~\ref{fig:energyFlux}, middle row) have structure corresponding to the RT instability bubble and spike.
The ideal terms are concentrated within the bubble and spike, with the negative and positive regions in the spike and bubble, respectively.
However the non-ideal terms are concentrated in the lower collisionality low-density fluid.
The heat flux, $q_y$, similar to $n_N/n$, reaches a global maximum at the center of the spike, with a local maximum at the tip of the bubble.
Whereas the stress term, $u_i\Pi_{iy}$, reaches its maximum magnitude in the bulk of the bubble rather than the tip, where it also flips sign.
This is likely due to dominance of tangental stress terms in the lower region, compared to dominant positive vertical flux at the tip of the bubble.
Case 3 (Figure~\ref{fig:energyFlux}, bottom row) shows similar characteristics to both cases 1 and 2, where the ideal terms are largely concentrated in the RT instability structures and the more collisional upper region, while the non-ideal terms are concentrated in the less collisional lower region.
The vertical asymmetry in bubble and spike formation is easily seen through the energy-flux, as the values reached by the ideal terms within the bubble are substantially greater than those reached in the spike.
Differences between the pressure term, $5/2u_y p$, and the inertial term, $1/2mnu_y \boldsymbol{u}^2$, at the spike position show energy tends to go into compression rather than downward movement of the gas.
Relative importance of the ideal and non-ideal terms can be quantified by taking the ratio of the averages of the absolute values of the non-ideal terms to the ideal terms,
\begin{equation}
    \text{ratio} = \frac{\text{avg}(\left|q_k + u_i\Pi_{\mathit{ik}}\right|)}{\text{avg}(\left|\frac{5}{2}u_kp + \frac{1}{2}mnu_k\boldsymbol{u}^2\right|)}.
\end{equation}
For case 1, the low collisional case dominated by kinetic non-ideal behavior, the ratio is approximately 1.952.
Cases 2 and 3 are expected to have more relevant ideal term dynamics related to instability growth, and the ratios for those cases are 0.037 and 0.873, respectively.
Case 3, while still dominated by the ideal terms, has a higher ratio than case 2 due to the lower average collisionality.

\begin{figure}[h!]
    \centering
    \includegraphics[width=0.8\linewidth]{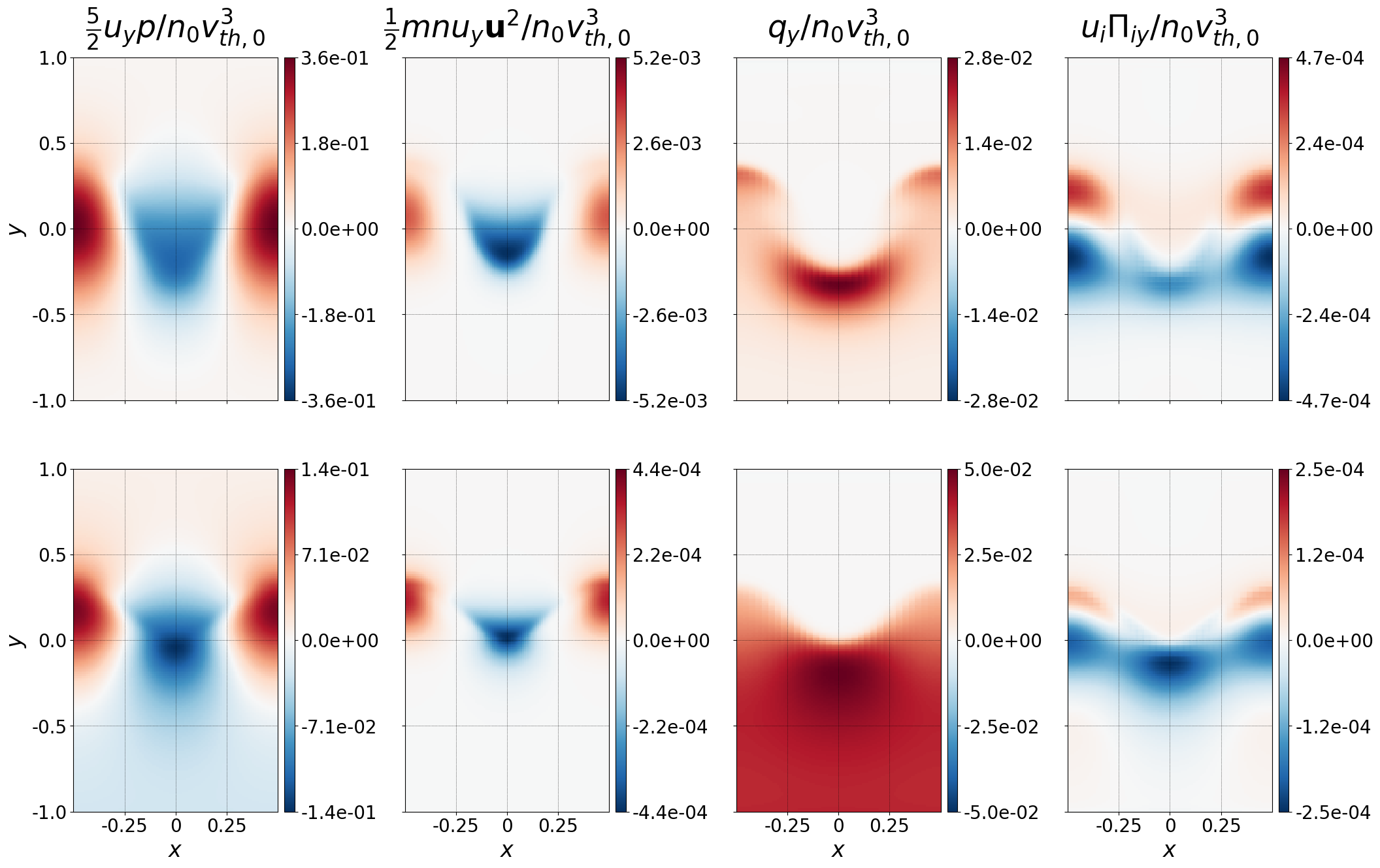}
    \caption{Terms in the expanded particle energy-flux, Eq. 14, for cases 2 (top) and 3 (bottom) with Atwood number $A=2/3$. Similar to the lower Atwood number cases, the ideal terms (left two columns) are concentrated in the bubble and spike, while the non-ideal terms (right two terms) are concentrated in the low-collisionality regions with extrema around the interface.}\label{fig:energyFlux23}
\end{figure}

A similar analysis can be applied to the higher Atwood number cases, shown in Figure~\ref{fig:energyFlux23}.
In case 2 (Figure~\ref{fig:energyFlux23}, top row), each term of the expanded particle energy flux is of the same order of magnitude as the lower $A$ cases.
However, one notable change is the increase in magnitude of the non-ideal terms by approximately a factor of 2, similar to the increase present in $n_N/n$.
The ratio of ideal to non-ideal terms for this case is 17.81, a similar magnitude to the lower Atwood number case and still dominated by ideal behavior.

Case 3 (Figure~\ref{fig:energyFlux23}, bottom row) has a ratio of 1.58 and shows substantial differences to the lower $A$ case, primarily centered around the spike.
Both ideal terms increase in magnitude as $A$ increases, and the maxima reached in the spike are comparable to those in the bubble, whereas the lower $A$ case shows much smaller magnitudes in each term in the spike relative to the bubble.
This is especially true in the inertial term $mnu_y\boldsymbol{u}^2/2$, where the ratio of peak magnitude in the bubble to the spike is 0.789 in the $A=2/3$ case, compared to 3.389 in the $A=1/3$ case.
Similar to case 2, the non-ideal terms both approximately double in magnitude, contributing to the similar increase in $n_N/n$.
The non-ideal terms of the particle energy-flux are purely kinetic effects, and spatial variation of collision frequency contributes further to non-ideal behavior by varying effects such as diffusion and viscosity.
Specifically, the presence of these non-ideal terms in the expansion of the third moment of $f$ imply the presence of skewness in the distribution function that would not be captured by fluid models.

\section{Summary}
\label{sec:summary}
Simulations of the single-mode Rayleigh-Taylor (RT) instability in 2X3V (2 spatial dimensions, 3 velocity space dimensions) are successfully conducted using a continuum-kinetic model implemented within \texttt{Gkeyll} with a more realistic, spatially-varying collision frequency.
Three cases are explored with an Atwood number $A=1/3$, enumerated in Table 1, covering regimes previously studied with spatially constant collisionality.
Case 1 exhibits no Rayleigh-Taylor instability growth, despite the upper region being in the intermediate collisional regime that shows instability growth for previous constant collision frequency simulations.
Case 2 varies between intermediate and highly-collisional regimes, and the RT instability growth is the most fluid-like of the three cases.
The fully-varying case 3 that covers low and high collisional regimes exhibits limited RT instability growth with a high degree of diffusion in the low collisional region that results in an upward movement of the interface.

Variations from Maxwellian distribution function can be quantified by taking the zeroth velocity moment of the difference between the local distribution function and a Maxwellian constructed from its moments to calculate the non-Maxwellian density, $n_N$.
In all cases, the normalized non-Maxwellian density, $n_N/n$, is localized to regions of low collisionality.
Additionally, there are extrema in $n_N/n$ around the interface in cases 2 and 3, where the instability develops, likely related to compressibility effects.
Higher relative magnitudes in $n_N/n$ at the spikes than the bubbles of cases 2 and 3 are connected to the damping of the downward growth of the spike.

Increasing the Atwood number from 1/3 to 2/3 for cases 2 and 3 yields larger instability amplitude stemming from greater downward movement of the spike into the low density region.
Diffusion in the low collisionality region also moves the interface upwards to a lesser degree than the $A=1/3$ cases.
Non-Maxwellian density follows the same trends as the lower Atwood number cases, though the magnitudes are higher in the higher Atwood number cases.
This is likely due to the lower collision frequency and globally lower density and higher temperature in the latter cases and the proportionality of kinematic viscosity and diffusion coefficient to thermal velocity.

Growth rates calculated from tracking the instability amplitude generally match the modified theoretical growth rates, $\gamma_0$, that include the effects of diffusion and viscosity.
Case 2, which has the highest average collision frequency, agrees well with $\gamma_0$ for each Atwood number.
However, case 3, which has the greatest collisional variation, matches the theoretical prediction much more closely for $A=2/3$ than for $A=1/3$, implying a sensitivity to Atwood number for the highly collisionally varying cases.  

As the instability evolves, the width of the interface changes in time due to free-streaming particle diffusion.
In a traditional inviscid fluid model, numerical diffusion, rather than a physical mechanism, is the primary source of interface widening, so the continuum-kinetic model is expected to capture the interface width evolution more accurately.
Further, there is a difference between the interface width at the center of the spike and the center of the bubble.
The lower, less-collisional region is the region where most diffusion occurs in the domain, so as the bubble moves upward, the lower end of the interface diffuses away downward, spreading the interface.
At the center of the spike, the upper end of the interface moves downward as the instability grows, limiting the spread of the interface due to diffusion.
For both cases, the higher $A$ runs show greater widening of the interface than the corresponding lower $A$ runs.
Case 2 for both Atwood numbers exhibits similar interface widening to the spatially constant intermediate collisionality $\mathit{Kn}=0.01$ case.
With $A=2/3$, case 3 most closely resembles case 1 with $A=1/3$, which is effectively pure diffusion, while case 3 with $A=1/3$ is similar to case 2 and $A=2/3$, further showing the sensitivity of case 3 to Atwood number.

An expansion of the particle energy-flux is utilized to highlight the presence of kinetic effects and the presence of non-Maxwellian dynamics.
While the dominant components of the energy-flux come from the ideal terms arising from the Maxwellian components of the distribution function, there are concentrations of non-ideal terms that only appear when using a distribution function that can deviate from Maxwellian.
In general, the non-ideal components of the energy-flux are primarily present in regions of low collisionality, i.e. the lower region, and are inversely proportional in magnitude to the collision frequency.
The global extrema of each term occur around the interface when the RT instability grows, with the maximum magnitude occurring in the spike, where corresponding density compression and maxima in $n_N/n$ are also present.
Non-ideal terms of the particle energy-flux, the laboratory-frame third moment of the distribution function, implies the presence of skewness away from Maxwellian, which would not be inherently captured by fluid models.
Additionally, simple high collisional limit models of viscosity do not account for intermediate and low collisionality effects.

While fluid models generally are well-suited to study the RT instability, collisional effects can be relevant in astrophysical and laboratory plasmas.
Previous continuum-kinetic simulations with constant collision frequency demonstrate the existence of intermediate regimes that are not accessible to traditional fluid models.
Simulations with spatially-varying collision frequency offer further variations from the high-collisionality fluid limit, emphasizing the necessity of kinetic models to capture dynamics that would be missed with traditional fluid models.
Further simulations are necessary to expand a kinetic study into two-component plasmas with self-consistent electromagnetic fields both with and without collisions and externally applied fields.
Introducing such complications adds to the number of relevant spatial and temporal scales in the system, increasing the likelihood that kinetic physics will be relevant.

\section{Acknowledgements}
This work was supported by the National Science Foundation CAREER award under grant number PHY-1847905.
J. Juno was supported by the U.S. Department of Energy under Contract No. DE-AC02-09CH1146 via an LDRD grant.
The authors acknowledge Advanced Research Computing at Virginia Tech for providing computational resources and technical support that have contributed to the results reported within this paper (\url{http://www.arc.vt.edu}). 

\section{Getting Gkeyll and reproducing results}
Readers may reproduce our results and also use Gkeyll for their applications. The code and input files used here are available online. 
Full installation instructions for
Gkeyll are provided on the Gkeyll website \citep{gkylDocs}. 
The code can be installed on Unix-like operating systems (including Mac OS and Windows using the Windows Subsystem for Linux) either by installing the pre-built binaries using the conda package manager (\url{https://www.anaconda.com}) or building the code via sources.

\newpage
\bibliography{ref}
\end{document}